\RequirePackage[final]{graphicx}
\documentclass[aps, superscriptaddress, prl, twocolumn, 10pt,floatfix]{revtex4-2}
\usepackage[utf8]{inputenc}
\usepackage{amsmath}
\usepackage{amssymb}
\usepackage{amsfonts}
\usepackage{mathtools}
\usepackage{fixme}
\usepackage[usenames,dvipsnames]{color}
\usepackage{braket}
\usepackage{dsfont}
\definecolor{columbiablue}{RGB}{0,114,206}
\usepackage[colorlinks,allcolors=columbiablue]{hyperref}

\def\bk{{\bf k}}

\def\bi{{\bf i}}
\def\bj{{\bf j}}

\def\nn{\nonumber}

\def\Ham{{ \hat{H} }}

\begin{document}
\title{Magnetic domains stabilized by symmetry-protected zero modes}
\author{Pavel Kos}
\affiliation{Max Planck Institute of Quantum Optics, Hans-Kopfermann-Stra{\ss}e 1, D-85748 Garching, Germany}
\affiliation{Department of Physics, Faculty of Mathematics and Physics, University of Ljubljana, Jadranska 19, SI-1000 Ljubljana, Slovenia}
\author{Dominik S. Wild}
\affiliation{Max Planck Institute of Quantum Optics, Hans-Kopfermann-Str. 1, D-85748 Garching, Germany}
\affiliation{Google Quantum AI, D-80636 Munich, Germany}
\author{Kristian Knakkergaard Nielsen}
\affiliation{Niels Bohr Institute, University of Copenhagen, Jagtvej 128, DK-2200 Copenhagen, Denmark}
\date{\today}

\begin{abstract}
Understanding mechanisms for the breakdown of thermalization in closed quantum systems is a central problem in quantum many-body physics. We demonstrate strong nonergodic behavior in the XX model on coupled chains, where domain-wall initial states retain an inhomogeneous magnetization profile for arbitrarily long times. We find that this effect arises due to zero modes protected by chiral symmetry. Using an analysis based on the Lanczos algorithm, we identify evidence for a localization transition in the thermodynamic limit at a critical coupling between the chains. We further show that antiferromagnetic defects in the initial state and symmetry-breaking perturbations restore slow thermalization, whereas it remains robust for symmetry-conserving perturbations. These results establish that degenerate, symmetry-protected subspaces can give rise to thermodynamically stable nonergodic dynamics in experimentally accessible quantum systems.
\end{abstract}

\maketitle

\paragraph{Introduction.--}
Central to our current understanding of quantum many-body systems, the eigenstate thermalization hypothesis (ETH) \cite{Deutsch1991,Srednicki1994,Garrison2018} explains how closed quantum systems thermalize: at long times local observables are describable by a thermal ensemble with a temperature set by the average energy of the initial state. 
Four main mechanisms that challenge this paradigm have been identified: integrability \cite{Rigol2008,Rigol2009,Alessio2016}, many-body localization (MBL) \cite{Nandkishore2015,Imbrie2016,Bloch2017,Panda2019,Abanin2021,Sierant2022,Long2023,Evers2023,DeRoeck2016,Sels2021,Morningstar2022,Suntajs2020}, Hilbert-space fragmentation (HSF) \cite{Sala2020,Moudgalya2022b,Brighi_2023, Adler2024,Zhao2025,Yang2025a,orlov2025loopchargesfragmentationpairwise} and quantum many-body scars (QMBSs) \cite{Bernien2017,Turner2018,Bluvstein2021,Maskara2021,Moudgalya2022,Desaules2022,SY_Zhang2023,Wang2024,Kerschbaumer2025}. MBL and HSF arising from disorder and exponentially many disconnected subspaces, respectively, are scenarios by which ETH can be strongly broken, in that the entire Hilbert space may change character. Local observables are, thus, no longer a smooth function of energy and thermalization can only be established in a limited sense. Integrability leads to the so-called generalized Gibbs ensemble, by which a modified version of thermalization remains. Finally, many-body scars are often characterized as a rare instance in which a small subset of initial states resist thermalization due to their overlap with rare nonthermal eigenstates. We note, however, that the precise definition of QMBS remains under active debate \cite{Chandran2023,Moudgalya2024,Yao2025}.

\begin{figure}[t!]
    \centering
    \includegraphics[width=1\linewidth]{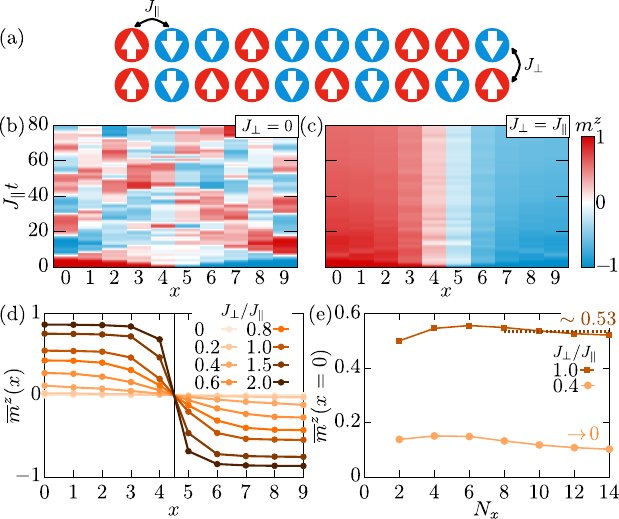}
    \vspace{-0.75cm}
    \caption{(a)~We consider lattices of $N = N_x\times N_y$ ($N_y = 2$ here) spin-$1/2$ particles interacting with direction-dependent nearest neighbor couplings $J_\parallel, J_\perp$. (b),(c) Rung magnetization dynamics  $m^z(x,t) = \sum_y \braket{\hat{S}^z_{(x,y)}(t)}$ from an initial state with a single ferromagnetic domain wall. (b)~Decoupled chains ($J_\perp = 0$) map to free fermions and show undamped ballistic spin motion. (c)~Nonzero $J_\perp$ leads to non-thermalizing nonhomogeneous magnetization profiles. (d)~Longtime average rung magnetization, $\overline{m}^z(x)$, showing  stronger domains as $J_\perp / J_\parallel$ increases. (e) Longtime average of the magnetization at the left edge, $\overline{m}^z(x = 0)$, vs even $N_x$ with $N_y = 2$. The initial state has total magnetization $S^z = 0$ with two equally sized ferromagnetic domains. For large even $N_x$, $\overline{m}^z(x = 0)$ approaches a nonzero constant if $J_\perp \gtrsim 0.5 J_\parallel$, as indicated with the suggested plateau number, following from Secs. \emph{Stability of nonergodic dynamics} and \emph{Localization transition}.}
    \label{fig:Fig1}
    \vspace{-0.75cm}
\end{figure}

In this Letter, we unveil nonergodic behavior induced by symmetry-protected zero modes. We show that this leads to \emph{thermodynamically stable} magnetic domains in the XX model on two-dimensional square lattices of size $N = N_x \times N_y$. Indeed, while a single chain XX model features free fermionic excitations with \emph{ballistic transport} of the spins, coupling two chains leads to the appearance of persistent domain walls for arbitrarily long times. By analyzing spin ladders up to sizes of $N = N_x \times N_y = 14\times 2$, we find a \emph{localized regime} for sufficiently strong inter-leg couplings, in which domain walls remain robust in the thermodynamic limit. We identify that this striking effect is sustained by exponentially many zero modes, eigenstates at energy $E = 0$, which are protected by chiral symmetry \cite{Weinberg_1981,Witten1982,Schecter2018}. While this localization effect is robust to symmetry-conserving perturbations, symmetry breaking and antiferromagnetic defects in the initial state reestablish thermalization.  Moreover, the effect vanishes when $N_y$ is odd whereas it persists for even values greater than $2$, albeit more weakly. In this manner, we show that the appearance of an exact zero-mode subspace has profound physical implications in systems ripe for experimental investigations using contemporary quantum simulators \cite{Greiner2002,Roushan2017,Zhiguang2019,Zhu2022,Barredo2015}. We emphasize that the observed nonergodicity appears even though the XX model on rectangular lattices is non-integrable and coexists with otherwise ergodic high-temperature diffusive behavior \cite{Znidaric2013,Steinigeweg2014,Karrasch2015,Kloss2018,Rakovszky2022}, as well as scrambling dynamics from antiferromagnetic initial states \cite{Zhu2022}. In the broad sense of rare nonthermal eigenstates, these zero modes may themselves be regarded as QMBSs; given the ongoing debate over the precise definition of the term, however, we refer to them throughout as symmetry-protected zero modes.

\paragraph{Model.--} We consider spin-$1/2$ particles in a $N_x\times N_y$ square lattice described by the XX model, 
\begin{align}
\Ham &= \sum_{\substack{\lambda = \parallel,\perp \\ \braket{\bi,\bj}_\lambda}} \! J_\lambda \left[\hat{S}^x_\bi\hat{S}^x_\bj + \hat{S}^y_\bi\hat{S}^y_\bj\right],
\label{eq.H}
\end{align}
in which $J_\parallel, J_\perp$ denote the nearest neighbor spin coupling along the $x$- and $y$-directions [see Fig.~\ref{fig:Fig1}(a)]. $\hat{S}^\alpha_\bi = \hat{\sigma}^\alpha_\bi/2$ are the spin-1/2 spin operators acting on site $\bi=(x,y)$. Since $\hat{S}^z = \sum_\bi \hat{S}^z_\bi$ commutes with the Hamiltonian it is conserved. We note that the XX model can be obtained from the Bose-Hubbard model at strong on-site interactions $U\gg J_\lambda$ readily implementable with ultracold atoms in optical lattices \cite{Jaksch1998,Greiner2002} and superconducting processors \cite{Roushan2017,Zhiguang2019,Zhu2022}, as well as dipolar molecular and Rydberg atomic arrays \cite{Gorshkov2011a,Hazzard2013,Barredo2015,Yang2025b,Ding2026}. Our predictions are, thus, readily testable in quantum simulation experiments.

\begin{figure}[b!]
    \centering
    \includegraphics[width=1\linewidth]{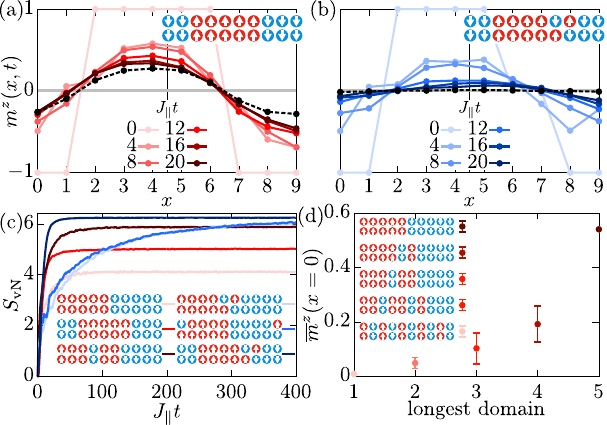}
    \vspace{-0.75cm}
    \caption{\textbf{Stability of nonergodic dynamics.} (a), (b)~Magnetization profile, $m^z(x,t)$, at indicated times $J_\parallel t$ for shown initial product states (insets) compared to their temporal average (dashed black), as well as the ETH expectation (gray), revealing nonthermal behavior in (a) and thermalization in (b). (c)~Bipartite von Neumann entanglement entropy dynamics, $S_{\rm vN}$, for indicated initial product states. Thermalizing states (blue) approach a similar value, $S_{\rm vN} > 6$. Non-thermalizing states (red) leads to lower $S_{\rm vN}$ dependent on the initial states. (d)~Time-averaged magnetization at $x = 0$ vs length of the longest spin-$\uparrow$ domain in rung-ferromagnetic initial states (legend shows exemplary initial states). Values are obtained by averaging over all possible initial states with the longest domain starting at $x = 0$. Error bars show the associated standard deviation. Parameters: $J_\perp = J_\parallel$, $N = 10\times 2$.}
    \label{fig:Fig2}
    \vspace{-0.75cm}
\end{figure}

\paragraph{Stable domains in the XX limit.--}
To exemplify the emergent stability of magnetic domains, we initially focus on a $N = N_x\times 2$ spin ladder at zero magnetization, $N_\uparrow = N_\downarrow = N/2$. We assume that the initial state is a perfect left-right domain wall state: $\ket{\Psi(0)} = \ket{\Uparrow\cdots\Uparrow \Downarrow\cdots\Downarrow}$, with $\ket{\Uparrow} = \ket{\substack{\uparrow\\\uparrow}}$ and $\ket{\Downarrow} = \ket{\substack{\downarrow\\\downarrow}}$, whereby all spin-$\uparrow$ particles are initially on the left half of the ladder \footnote{When $N_x$ is odd and at zero total magnetization, we need to insert at least a single antiferromagnetic rung $\ket{\substack{\uparrow\\\downarrow}}$ between the ferromagnetic domains. This leads to delocalization as demonstrated in Fig.~\ref{fig:Fig2}.}. Figures \ref{fig:Fig1}(b) and \ref{fig:Fig1}(c) show the associated dynamics of the rung magnetization,
\begin{equation} \label{eq.rung_magnetization}
m^z(x,t) = \sum_{y}\bra{\Psi(t)}\hat{S}^z_{(x,y)}\ket{\Psi(t)}, 
\end{equation} 
in the time-evolved state $\ket{\Psi(t)} = \exp(-i\Ham t) \ket{\Psi(0)}$ for $J_\perp = 0$ and $J_\perp = J_\parallel$, respectively. When the chains are disconnected, $J_\perp = 0$, a Jordan-Wigner transformation \cite{Jordan1928} allows them to be described by free fermions. The associated dynamics seen in Fig.~\ref{fig:Fig1}(a) is the resulting ballistic motion of the spins. We note that the temporal average of these oscillations, $\overline{m}^z(x) = \lim_{T \to \infty} T^{-1} \int_0^T dt \, m^z(x,t)$, is exactly zero in this case. 

However, when the coupling between the chains are turned on something remarkable happens [Fig.~\ref{fig:Fig1}(c)]: the magnetization now settles into a nonhomogeneous profile. Fig.~\ref{fig:Fig1}(d) shows that this effect is already significant for $J_\perp \gtrsim 0.6$. We note that this behavior is in complete contradiction with what one would expect from ETH. Indeed, since the average energy of the state is zero, $\bra{\Psi(t)} \Ham \ket{\Psi(t)} = 0$, the effective temperature is infinitely high, $T_E = \infty$. In the associated infinite temperature thermal ensemble, the spin-$\uparrow$ particles are, thus, equally distributed with $m^z_{\rm th}(x) = 0$. Figure \ref{fig:Fig1}(e), finally, shows the time-averaged rung magnetization on the left-most rung, $\overline{m}^z(x = 0)$. This demonstrates that for even $N_\uparrow = N_x$ and large enough $J_\perp$, the time-averaged magnetization $\overline{m}^z(x = 0)$ approaches a nonzero value for $N_x \gg 1$ \footnote{For $N_x \leq 8$, we use exact diagonalization for the longtime average. For larger $N_x$, we compute the exact state dynamics with the ``expv'' library in Julia to very long times ($J_\parallel t_{\rm max} = 10^3$) and compute the averages numerically.}. For smaller $J_\perp$, we instead find that the time-averaged magnetization will eventually vanish.

\paragraph{Stability of nonergodic dynamics.--} We now analyze how different initial states lead to distinct dynamics. Figures \ref{fig:Fig2}(a) and \ref{fig:Fig2}(b) show $m^z(x,t)$ for two slightly different initial states. These examples illustrate that for the domain wall to be stable at long times, it has to start out in a state with perfect ferromagnetic correlations on each rung of the spin ladder: $C_{\rm ZZ}(x) = 4\braket{S^z_{(x,0)}S^z_{(x,1)}} = 1$ [as in Fig.~\ref{fig:Fig2}(a)]. We will refer to these states as \emph{rung-ferromagnetic} states. On the other hand, in Fig.~\ref{fig:Fig2}(b) the initial state has two \emph{anti}ferromagnetic rungs, leading to thermalization.  

The thermalization behavior may also be tracked by the bipartite von Neumann entanglement entropy, $S_{\rm vN}(t) = -{\rm tr}[\hat{\rho}_{\mathcal{L}}(t)\ln \hat{\rho}_{\mathcal{L}}(t)]$, where $\hat{\rho}_{\mathcal{L}}(t) = {\rm tr}_{\mathcal{R}}[\ket{\Psi(t)}\bra{\Psi(t)}]$ is the reduced density operator of the left half ($\mathcal{L}$) of the spin ladder, obtained by tracing out the spins on the right half ($\mathcal{R}$). The resulting entanglement entropy dynamics for differing initial states are shown in Fig.~\ref{fig:Fig2}(c). This clearly demonstrates that in the thermalizing cases the entanglement entropy approaches a universal value close to the Page value, $S_{\rm Page} \simeq (N \cdot \ln 2 - 1)/2$ \cite{Page1993,Foong1994}. On the contrary, for nonergodic initial states, the entanglement entropy settles at values dependent on the initial state significantly below $S_{\rm Page}$. Figure \ref{fig:Fig2}(d) finally shows that the longtime stability of the domains is retained even as the magnetic domains are split up, albeit with a smaller magnetic plateau for shorter domains.

\paragraph{Localization transition.--} We now use the Lanczos algorithm \cite{Lanczos1950,Paige1972} to argue that the overlap of the domain wall initial states remains finite in the thermodynamic limit if the coupling between chains is strong enough. To this end, we compute the Hamiltonian in the Krylov basis $\{\ket{K_n}\}$  starting from an initial domain wall state $\ket{K_1} = \ket{\Uparrow \cdots \Uparrow \Downarrow \cdots \Downarrow}$ (see End Matter), resulting in the effective one-dimensional tight-binding model
\begin{align}
H_{\rm eff}[\ket{K_1}] = \sum_j \beta_{j + 1}\Big[\ket{K_{j + 1}} \bra{K_j} + \ket{K_j} \bra{K_{j + 1}} \Big].
\label{eq.H_lanczos}
\end{align}
While the Lanczos method generally results in on-site terms $\alpha_j \ket{K_j}\bra{K_j}$, these vanish for the XX model in Eq.~\eqref{eq.H} due to a chiral symmetry $\hat{\mathcal{S}} = \prod_{\bi \in \mathcal{E}} (2\hat{S}^z_\bi)$, satisfying $\{\hat{\mathcal{S}},\Ham\} = 0$. Here, $\mathcal{E} = \{\bi = (x,y)\!: x + y \; {\rm even}\}$ denotes the sublattice of all the even sites. From Eq.~\eqref{eq.H_lanczos}, a unique zero mode follows: $\ket{\psi_0} = \sum_{j} c_j \ket{K_j}$, with nonzero coefficients for the odd indices, $c_{2n + 1} = \prod_{l = 1}^n(-\beta_{2l}/\beta_{2l + 1}) \times c_1$. The Lanczos coefficients thus determine the overlap of this zero mode with the initial state. 

\begin{figure}[t!]
    \centering
    \includegraphics[width=1\linewidth]{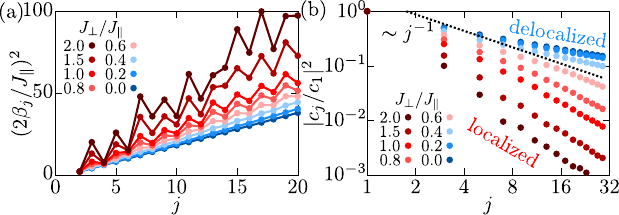}
    \vspace{-0.75cm}
    \caption{\textbf{Thermodynamic stability.} (a)~Lanczos coefficients for indicated values of $J_\perp/J_\parallel$ starting from the initial state $\ket{K_1} = \ket{\Uparrow \cdots \Uparrow \Downarrow \cdots \Downarrow}$. As $J_\perp$ increases, pronounced oscillations appear. (b)~The associated coefficients $|c_j|^2$ in the zero-energy eigenstate $\ket{\psi_0} = \sum_{j} c_j \ket{K_j}$ features a clear crossover behavior. They decay slower than $j^{-1}$ for $J_\perp \lesssim 0.5 J_\parallel$, and \emph{faster} than $j^{-1}$ for $J_\perp \gtrsim 0.5 J_\parallel$. This marks a \emph{transition} from a macroscopically delocalized regime (blue data) to a \emph{localized} regime (red data) as $J_\perp$ is increased. We use a system size of $N = 14\times 2$. The finite size effects are limited (see Supplemental Material~\cite{SM}).}
    \label{fig:Fig3}
    \vspace{-0.5cm}
\end{figure}

By computing the Lanczos coefficients numerically, we find that in addition to an overall linear growth of $(\beta_j)^2$, they display pronounced oscillations depending on whether $j$ is even or odd. This results in faster power-law decay of $|c_j|^2\sim j^{-\gamma}$ for increasing $J_\perp$ [Fig.~\ref{fig:Fig3}]. Consequently, the exponent goes from $\gamma \leq 1$ for $J_\perp \leq J_\perp^c \simeq 0.5 J_\parallel$ to $\gamma > 1$ for $J_\perp > J_\perp^c$. When $J_\perp \leq J_\perp^c$, the sum $\sum_{j=1}^n |c_j/c_1|^2 \sim n^{1 - \gamma}$ \emph{does not} converge for $n\to \infty$ \footnote{In the borderline case of $J_\perp = J_\perp^c$, the sum \emph{diverges} as $\ln(n)$}. As a result, the overlap with the initial state $\ket{K_1}$ vanishes as $N_x \to \infty$. On the other hand, when $J_\perp > J_\perp^c$, the sum \emph{converges}, and the zero mode retains a nonzero overlap with the initial state as $N_x \to \infty$.

The appearance of this \emph{localization transition} can qualitatively be understood as follows. We approximate $(\beta_{j})^2 = a \, j + b_{o/e}$, where $b_o \geq b_e$ applies to odd and even $j$, respectively. This works well for $J_\perp \leq J_\parallel$. For $N_y$ uncoupled legs ($J_\perp = 0$), we indeed analytically find $b_o = b_e$ and $(2\beta_{j}/J_\parallel)^2 = N_y (j - 1)$ \cite{SM}. As a result of this double linear structure, the coefficients feature power-law decay $\left|c_{2n + 1}/c_1\right|^2 \propto n^{-\gamma}$ with exponent
\begin{equation}
\gamma = \frac{1}{2}\left[1 + \frac{b_o-b_e}{a}\right].
\label{eq.gamma_model}
\end{equation}
Hence, more oscillatory $\beta_j$ (increasing $b_o - b_e > 0$) leads to faster power-law decay. The exponent $\gamma$, thus, eventually crosses $\gamma_c = 1$ for large enough $J_\perp$. 

The oscillations in $\beta_j$, therefore, are an \emph{essential} feature of the localization of the magnetic domains (together with the vanishing of the diagonal elements $\alpha_j$), akin to recent analyses of Anderson localization \cite{Peacock2026}. Note that the Lanczos coefficients $\beta_{j}$ are proportional to $\sim \Vert \Ham\ket{K_{j-1}} \Vert^2$ such that $\beta_{j}$ relates to the mobility of the state $\ket{K_{j-1}}$. Hence, the localization can be seen as an intriguing consequence of \emph{structured mobility}, by which every odd state $\ket{K_{2j-1}}$, associated with $\beta_{2j}$, is comparatively \emph{less mobile} than the next even state $\ket{K_{2j}}$, associated with $\beta_{2j + 1}$. Indeed, for the odd states, \emph{rung-ferromagnetic} configurations tend to appear. This has suppressed mobility because the rungs contain identical spins, rendering them inactive under $\Ham$. Conversely, if the initial state has a single \emph{anti}ferromagnetic rung, the behavior flips, $b_o < b_e$, and the formed zero mode is delocalized (see End Matter).

\begin{figure}[t!]
    \centering
    \includegraphics[width=1\linewidth]{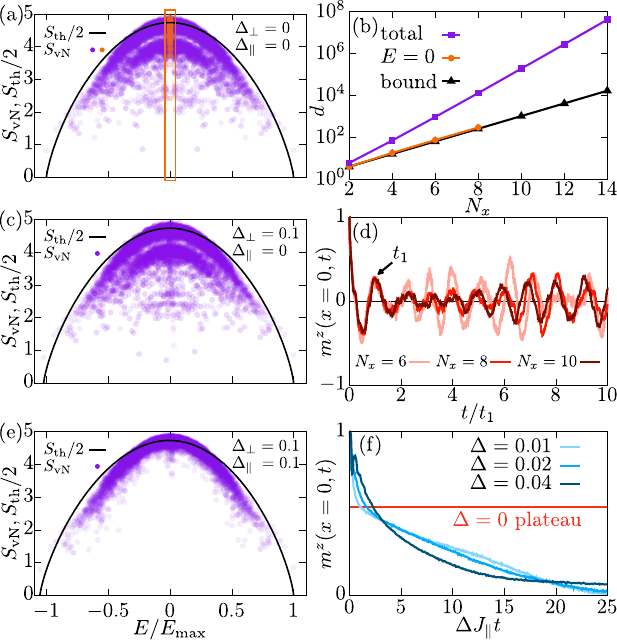}
    \vspace{-0.75cm}
    \caption{\textbf{ZZ interactions.} (a),(c),(e)~Bipartite von Neumann entropy $S_{\rm vN}(E)$ of the energy eigenstates in the $S^z = 0$ sector for $8\times 2$ spin ladders with indicated ZZ interaction strengths, $\Delta_\perp,\Delta_\parallel$. This is compared to half the canonical thermal entropy $S_{\rm th}(\beta_E)/2$, where $\beta_E$ is chosen such that the average thermal energy equals $E$. (a)~Many states appear exactly at $E = 0$ (orange dots within orange box). (b)~Degeneracy of these states as a function of $N_x$ (orange dots) alongside the total $S^z = 0$ magnetization sector dimension, $d_{\rm tot} = \binom{N}{N/2}$ (purple squares), and the exponential lower bound $2^{N/2}$ [Eq.~\eqref{eq.tr_C}] from chiral symmetry (black triangles). (c)~ZZ rung couplings remove the $E = 0$ degeneracy, but states with low entanglement akin to quantum many-body scars persist. (d)~The magnetization dynamics features infinitely long-lived oscillations. The time axis is rescaled to the first revival ($J_\parallel t_1 \simeq 400, 625, 900$ for $N_x = 6, 8, 10$, respectively). (e)~The scarlike states disappear in the full XXZ model and $S_{\rm vN}$ is close to the thermal value expected from ETH. (f)~The resulting magnetization decays on a timescale $(\Delta J_\parallel)^{-1}$, $\Delta \equiv \Delta_\perp = \Delta_\parallel$ after passing the $\Delta = 0$ magnetization plateau (red line). In all panels, $J_\perp = J_\parallel$. } \label{fig:Fig4}
    \vspace{-0.5cm}
\end{figure}

\paragraph{Spectral analysis.--} To get an in-depth understanding of the longtime behavior, we use exact diagonalization to perform a full spectral decomposition up to system sizes of $N = 8\times 2$. We also compute the bipartite von Neumann entanglement entropy $S_{\rm vN}(E)$ for all eigenstates, $\ket{E}$, in the $S^z = 0$ magnetization sector. The results are shown in Figs.~\ref{fig:Fig4}(a), \ref{fig:Fig4}(c) and \ref{fig:Fig4}(e), where the latter two include ZZ spin couplings
\begin{align}
\Ham_{\rm ZZ} &= \!\!\sum_{\substack{\lambda = \parallel,\perp \\ \braket{\bi,\bj}_\lambda}} \!\!\! \Delta_\lambda J_\lambda \hat{S}^z_\bi\hat{S}^z_\bj,
\label{eq.H_ZZ}
\end{align}
as a perturbation to the XX model in Eq.~\eqref{eq.H} \footnote{The eigenstates are constructed such that they are also eigenstates of 3 additional commuting symmetries of the system: the point group reflection symmetries along the horizontal and vertical axes, $\hat{\sigma}_h, \hat{\sigma}_v$, as well as the total spin-flip symmetry $\hat{X} = \prod_\bi (2\hat{S}^x_\bi)$ present in the zero magnetization sector $S^z = 0$.}. We further include the canonical thermal entropy $S_{\rm th}(\beta) = \beta \braket{\Ham}_\beta + \ln(Z)$ of the Gibbs state $\hat{\rho}_\beta = \exp(-\beta \Ham)/Z$ for comparison. Here, $Z = {\rm tr}[\exp(-\beta \Ham)]$ is the partition function and $\braket{\Ham}_\beta = {\rm tr}[\hat{\rho}_\beta\Ham]$ is the average energy. The ETH expectation \cite{Garrison2018} is that the entanglement entropy is approximately $S_{\rm th}(\beta_E)/2$ \footnote{the factor of $1/2$ appears because the entanglement entropy is for a bipartition of the system into two equally sized subsystems}, where $\beta_E$ is the inverse temperature such that the average energy equals $E$: $\braket{\Ham}_{\beta_{E}} = E$. For the pure XX model [Fig.~\ref{fig:Fig4}(a)] and in the presence of ZZ rung couplings [Fig.~\ref{fig:Fig4}(c)], the entanglement entropy deviates strongly from the ETH expectation. Only in the full XXZ model [Fig.~\ref{fig:Fig4}(e)] does it appear to approach the ETH expectation. 

The deviation from ETH is accompanied by \emph{exponentially} many zero modes for \emph{even} $N_\uparrow = N_x$ as shown in Fig.~\ref{fig:Fig4}(b)~\footnote{For odd $N_\uparrow = N_x$, there, on the contrary, extremely few $E = 0$ states.}. It is the overlap with this degenerate $E=0$ subspace that allows the rung-ferromagnetic states to resist thermalization. Indeed, the temporal average of the local magnetization can be phrased in terms of the eigenstates $\ket{n}$ ($\Ham\ket{n} = E_n\ket{n}$) and the projection onto the zero mode subspace, $\ket{\Psi_P} \propto \hat{P}_0 \ket{\Psi(0)} = \sum_{n: E_n = 0} \ket{n}\braket{n|\Psi(0)}$, 
\begin{align}
&\bar{S}^z_\bi = \lim_{T \to \infty} \frac{1}{T} \int_0^T \bra{\Psi(t)} \hat{S}^z_\bi \ket{\Psi(t)} \nn \\
&= |c_P|^2  \bra{\Psi_P} \hat{S}^z_\bi \ket{\Psi_P} + \!\!\!\!\!\!\!\!\sum_{\substack{n,m: \\ E_n = E_m \neq 0}} \!\!\!\!\!\! c_n^* c_m  \bra{n} \hat{S}^z_\bi\ket{m},
\label{eq.local_magnetization_long_time}
\end{align}
with $c_n = \braket{n|\Psi(0)}$ and $c_P = \braket{\Psi_P|\Psi(0)}$ \footnote{Note that by construction $\ket{\Psi_P}$ is an eigenstate of the XX model with energy $E = 0$}. Numerically, we find that the contribution from the second term in Eq.~\eqref{eq.local_magnetization_long_time} vanishes \cite{SM}. As a result, the longtime stability of the magnetic domains, and in particular the actual magnetic profile at long times, comes from the overlap with the degenerate $E = 0$ subspace. Indeed, the employed Lanczos algorithm in \emph{Localization transition}, allows us to estimate that the domain wall product state with all spin-$\uparrow$ on the left for $J_\perp = J_\parallel$ has an overlap of $|\braket{\Psi_P|\Psi(0)}|^2 \simeq 0.53$ \cite{SM} in the thermodynamic limit, in excellent agreement with the observed bulk magnetization \footnote{ Since the projected zero mode is localized in Krylov space, the magnetization on the left edge of the projected state approaches $1$ as $N_x$ grows. Therefore, as $N_x$ grows, $\overline{m}^z(x = 0) = |c_P|^2 \sum_y \bra{\Psi_P} \hat{S}^z_{(x = 0, y)} \ket{\Psi_P} \to |c_P|^2$ is solely determined by the overlap with the projected zero mode.}  in Fig.~\ref{fig:Fig1}(e).

When the rung coupling $\Delta_\perp$ is turned on, the large zero mode degeneracy is lifted, but many low-entanglement states remain. As a result, the time evolution features extremely long lived revivals in the magnetization dynamics [Fig.~\ref{fig:Fig4}(d)] with a frequency that increases with the length of the system. This phenomenology bears close resemblance to the quantum many-body scars observed in Rydberg arrays \cite{Bernien2017,Bluvstein2021}, related to the PXP model \cite{Fendley2004,Lesanovsky2012}. Indeed, we identify that the oscillations are due to an unusually large overlap with only a few specific eigenstates. The model also hosts scar states \cite{SM} related to so-called cages \cite{Jonay2025,BenAmi2025,Tan2025} that persist even in the presence of ZZ rung couplings, $\Delta_\perp$. They have \emph{no significance}, however, for the nonergodic magnetic domains. Finally, in the full XXZ model with nonzero $\Delta \equiv \Delta_\perp = \Delta_\parallel$, the scar states disappear, and the magnetization thermalizes to $m^z(x) = 0$ on a timescale $(\Delta J_\parallel)^{-1}$ [Fig.~\ref{fig:Fig4}(f)].

\begin{figure}[b!]
    \centering
    \includegraphics[width=1\linewidth]{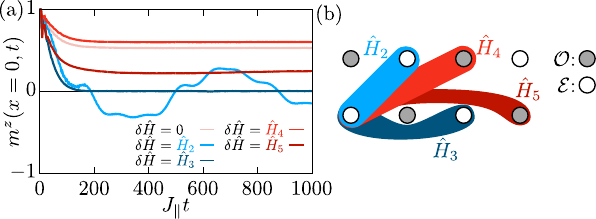}
    \vspace{-0.75cm}
    \caption{(a)~Magnetization dynamics for XX-type spin couplings $\Ham + \delta \Ham$ with 4 different perturbations $\delta\Ham = \Ham_n$  [see Eq.~\eqref{eq.H_n}]. (b)~$\Ham_2,\Ham_3$ couple sites on the same sublattice (both, e.g., on even sites $\mathcal{E}$), thereby breaking the chiral symmetry $\hat{C}$ and leading to unstable magnetic domains. By contrast, $\Ham_4,\Ham_5$ conserve the chiral symmetry and the domains remain robust. We use $J_\perp = J_\parallel$, $\delta J = 0.1 J_\parallel$ and $N = 12\times 2$.}
    \label{fig:Fig5}
    \vspace{-0.5cm}
\end{figure}

\paragraph{Symmetry protected zero modes.--} We now show that the nonergodic magnetic domains are sustained by a chiral symmetry. To this end, we consider $\hat{\mathcal{C}} = \hat{X}\hat{\mathcal{I}}\hat{\mathcal{S}}$, where $\hat{X} = \prod_\bi (2\hat{S}^x_\bi)$ is the total spin flip operator, $\hat{\mathcal{I}}$ is the spatial inversion operator and $\hat{\mathcal{S}} = \prod_{\bi\in\mathcal{E}} (2\hat{S}^z_\bi)$ is the sublattice operator considered after Eq.~\eqref{eq.H_lanczos}. Since $[\hat{X},\Ham] = [\hat{\mathcal{I}}, \Ham] = 0$ and $\{\hat{\mathcal{S}},\Ham\} = 0$, the formed operator is a chiral symmetry of the XX model in Eq.~\eqref{eq.H}: $\{\hat{\mathcal{C}}, \Ham\} = 0$. As we show in detail in Supplemental Material \cite{SM}, this chiral symmetry implies a \emph{lower bound} on the number of zero modes in the zero magnetization sector [shown in Fig.~\ref{fig:Fig4}(b)] in terms of the Witten index ${\rm tr}(\hat{\mathcal{C}})$ \cite{Weinberg_1981, Witten1982,Schecter2018},
\begin{equation}
d(E = 0) \geq |{\rm tr}(\hat{\mathcal{C}})| = \left\{\begin{matrix}
 2^{N/2}, & N_x, N_y \; {\rm even}.  \\
 0, & {\rm otherwise}.  \\
\end{matrix}\right.
\label{eq.tr_C}
\end{equation}
Hence, as long as $\hat{\mathcal{C}}$ is a chiral symmetry, there are at least $2^{N/2}$ zero modes for even $N_x, N_y$. This suggests that the discovered phenomenology is \emph{not} specific to the spin ladder. Indeed, we give numerical evidence that there is an even/odd effect in the number of legs, $N_y$, of the lattice, by which the magnetic domains are seen to remain for $N_y = 4$, but vanish for $N_y = 3$ \cite{SM}. Intriguingly, Eq.~\eqref{eq.tr_C} also suggests that the chiral symmetry stabilizes the nonergodic magnetic domains. To test this intuition, we compute the magnetization dynamics [see Fig.~\ref{fig:Fig5}(a)] for the total Hamiltonian $\Ham + \Ham_n$, for perturbations
\begin{align}
\Ham_n = \delta J\sum_{\braket{\bi,\bj}_n}  [ \hat{S}^x_\bi\hat{S}^x_\bj + \hat{S}^y_\bi\hat{S}^y_\bj],
\label{eq.H_n}
\end{align}
in which $\braket{\bi,\bj}_n$ denotes the $n$th nearest neighbor [illustrated in Fig.~\ref{fig:Fig5}(b)]. The perturbation respects the chiral symmetry when the $\bi,\bj$ pairs are on opposite sublattices and breaks it otherwise. We therefore expect that the nonergodic behavior survives only in the former case, which is confirmed by the numerical results shown in Fig.~\ref{fig:Fig5}(a). We note that the oscillations observed for $\Ham_2$ in Fig.~\ref{fig:Fig5}(a) are associated with the continued presence of quantum many-body scars \cite{SM}, similar to the ZZ rung coupling case in Fig.~\ref{fig:Fig4}(b). While the strong lower bound in Eq.~\eqref{eq.tr_C} only applies at zero magnetization, $S^z = 0$, we find a weaker, albeit still exponentially large, bound valid in all magnetization sectors based on the sublattice symmetry $\hat{\mathcal{S}}$ \cite{SM}, with a retained symmetry protected localization effect.

\paragraph{Conclusions and outlook.--} We have shown that symmetry-protected zero modes in the XX model give rise to an infinite lifetime of ferromagnetic domains. 
Our line of argument is as follows: (i) ferromagnetic domain-wall states display strongly nonergodic dynamics; (ii) a Lanczos analysis shows this stems from an exact $E=0$ eigenstate that localizes, in Krylov space, toward the initial state and retains a finite overlap with it above a critical coupling $J_\perp^c \approx 0.5\,J_\parallel$; (iii) this zero mode is one of exponentially many protected by a chiral symmetry; and (iv) the effect survives perturbations that respect the chiral symmetry and is destroyed by those that break it. Explicit symmetry breaking restores the relaxation of the domains in two perturbation-dependent ways: (1)~ complete melting of the domains as for the full XXZ model, or (2)~persistent oscillations at arbitrarily long times associated with the continued presence of cage-type quantum many-body scars \cite{SM,Tan2025,BenAmi2025,Jonay2025,Nicolau2026} as seen for the rung ZZ couplings and the next-nearest neighbor XX spin couplings. We note that the spin-$1/2$ XX ladder is generally  rich: beyond the symmetry-protected zero modes, it also exhibits Hilbert space fragmentation \cite{Znidaric2013} in subsectors \emph{orthogonal} to the ferromagnetic initial states we investigate; the remaining spectrum is chaotic in agreement with the Wigner surmise \cite{Znidaric2013}. 

Our analysis based on the Lanczos algorithm provides a powerful and intuitive framework, revealing an intriguing localization effect in terms of \emph{structured mobility}. Looking ahead, we believe that this methodology may provide a novel diagnostic tool to assess the stability of localization phenomena in the thermodynamic limit in a much larger range of scenarios. Moreover, the fact that the nonergodicity is stabilized by chiral symmetry suggests that similar behavior may be realized in many other quantum many-body systems --- in particular in settings where a different symmetry protects extensive degeneracies at nonzero energies. \\

\begin{acknowledgments}
The authors thank Nicolas Loizeau, Fan Yang, Klaus M{\o}lmer, Immanuel Bloch, Simon Karch, Marko Ljubotina, Cheryne Jonay and Marko \v{Z}nidari\v{c} for stimulating input. A special thanks to Nicolas Loizeau for guidance in using the Pauli strings implementation in Julia \cite{Loizeau2025}. K.K.N. acknowledges support from the Carlsberg Foundation through a Carlsberg Reintegration Fellowship (Grant No. CF24-1214). P.K. acknowledges funding from the European Union’s Horizon Europe research and innovation programme under the Marie Skłodowska-Curie grant agreement No. 101200860 (FAQ-QuantuM2D) and the Slovenian Research and Innovation Agency (ARIS) through grant J1-70061.
\end{acknowledgments} 

\emph{Data availability.--} The data files to produce all figures is openly available \cite{kristian_knakkergaard_nielsen_2026_19484074}. Source code will be shared upon reasonable request to K.K.N.

\bibliographystyle{apsrev4-2}
\bibliography{bibliography}

\appendix
\onecolumngrid

\section*{End Matter}

\twocolumngrid

In this End Matter, we describe: (A) the implemented Lanczos algorithm and why the diagonal terms in the obtained tri-diagonal matrix vanishes, (B) How this allows us to determine the asymptotic scaling of the coefficients of the $E = 0$ state in the Krylov basis within the double-linear model for the Lanczos coefficients, and (C) The Lanczos coefficients for other initial states than the left-right domain wall considered in the main text. \\

\paragraph{Appendix A: Lanczos algorithm and vanishing $\alpha_j$.--} The Lanczos algorithm follows the standard form \cite{Paige1972}
\begin{enumerate}
\item Start from some initial state $\ket{K_1}$.
\item Initial step:
\begin{enumerate}
    \item Let $\ket{w_1'} = \Ham \ket{K_1}$.
    \item Let $\alpha_1 = \braket{w_1'|v_1}$.
    \item Let $\ket{w_1} = \ket{w_1'} - \alpha_1 \ket{K_1}$.
\end{enumerate}
\item for $j = 2,\dots,m$:
\begin{enumerate}
    \item Let $\beta_j = [\braket{w_{j-1}|w_{j-1}}]^{1/2}$.
    \item If $\beta_j \neq 0$, let $\ket{K_j} = \ket{w_{j-1}}/\beta_j$.
    \item Let $\ket{w_j'} = \Ham \ket{K_j} - \beta_j \ket{K_{j-1}}$.
    \item Let $\alpha_j = \braket{w_j'|K_j}$.
    \item Let $\ket{w_j} = \ket{w_j'} - \alpha_j \ket{K_j}$.
\end{enumerate}
\end{enumerate}
This results in the tri-diagonal matrix in Eq.~\eqref{eq.H_lanczos}. Now, suppose that the initial state is an eigenstate of the sublattice symmetry: $\hat{\mathcal{S}}\ket{K_1} = c_1\ket{K_1}$, with $c_1 = \pm 1$. Our claim is that in this case $\alpha_j = 0$ for all $j$. We may prove this by inductively proving that  $\hat{\mathcal{S}}\ket{K_n} = (-1)^{n-1}c_1 \ket{K_n}$. 

To this end, note that by assumption $\hat{\mathcal{S}}\ket{K_1} = c_1 \ket{K_1}$, i.e. $\hat{\mathcal{S}}\ket{K_n} = (-1)^{n-1}c_1 \ket{K_n}$ holds for $n = 1$. Hence, the initial case in the inductive proof is fulfilled. Now, assume that $\hat{\mathcal{S}}\ket{K_j} = (-1)^{j-1}c_1 \ket{K_j}$ is true for $j\leq n$ for some $n\geq 1$. We need to show that it also holds for $n + 1$. This is the inductive step. To this end, note that $\ket{K_{n+1}} \propto \ket{w_n}$. So we just need to show it for $\ket{w_n}$. For $n\geq 2$,
\begin{align}
\hat{\mathcal{S}}\ket{w_n'} &= \hat{\mathcal{S}}\Ham\ket{K_n} - \beta_n \hat{\mathcal{S}} \ket{K_{n-1}} \nn \\
&= -\Ham\hat{\mathcal{S}}\ket{K_n} - \beta_n (-1)^{n-2}c_1\ket{K_{n-1}} \nn \\
&= (-1)^{n}c_1 \left[\Ham\ket{K_n} - \beta_n \ket{K_{n-1}}\right] \nn \\
&= (-1)^{n}c_1\ket{w_n'},
\end{align}
since $\hat{\mathcal{S}}\ket{K_n} = (-1)^{n-1}c_1\ket{K_n}$ by assumption. Since $\ket{w_n'},\ket{K_n}$ are both eigenvectors of the \emph{Hermitian} operator $\hat{\mathcal{S}}$ with \emph{different} eigenvalues, $\alpha_n = \braket{w_n'|v_n} = 0$. It follows that $\ket{w_n} = \ket{w_n'} - \alpha_n \ket{K_n} = \ket{w_n'}$. So $\hat{\mathcal{S}}\ket{w_n} = (-1)^nc_1 \ket{w_n}$, and therefore also $\hat{\mathcal{S}}\ket{K_n} = (-1)^n c_1 \ket{K_n}$. 

Note that it follows directly from this argument that $\alpha_n = 0$ for any $n$ as long as $\ket{K_1}$ is an eigenstate of $\hat{\mathcal{S}}$, proving our claim. Note finally that $\hat{\mathcal{S}} = \prod_{\bi\in{\rm E}} (2S^z_\bi)$ is a product of $\hat{S}^z_\bi$ operators. Therefore, all $\hat{S}^z$ basis states are eigenstates of $\hat{\mathcal{S}}$.

\paragraph{Appendix B: Asymptotic scaling of coefficients in the $E = 0$ state.--} 
Assume the double linear behavior of the Lanczos coefficients $(\beta_{2j})^2 = a\times 2j  + b_e, (\beta_{2j+1})^2 = a\times (2j + 1)  + b_o$. It follows that the coefficients of the $E = 0$ eigenstate behave as
\begin{align}
&\ln \left|\frac{c_{2n+1}}{c_1}\right|^2 = \sum_{j = 1}^n \ln \frac{(\beta_{2j})^2}{(\beta_{2j+1})^2} \nn \\
&= \ln A+\sum_{j = j_0}^n \left[\ln\left(1 + \frac{a^{-1} b_e}{2j}\right) - \ln\left(1 + \frac{1 + a^{-1} b_o}{2j}\right)\right] \nn \\
&\simeq \ln A' - \frac{1 + a^{-1}(b_o - b_e)}{2}\ln n,
\end{align}
where $A$, $A'$ are some constants, and we have used that $\ln (1 +x) \simeq x$ for $x \ll 1$ and $n, j_0 \gg a^{-1}b_e, a^{-1} b_o$. This shows that $|\frac{c_{2n+1}}{c_1}|^2 \to A' \cdot n^{-\gamma}$, with $\gamma = (1 + a^{-1}(b_o - b_e))/2$, as given in Eq.~\eqref{eq.gamma_model}. If $\gamma>1$ the coefficients decay fast enough to give a normalizable state. Otherwise the state is not normalizable, and we cannot get a simple state localized in the Krylov basis. 

\begin{figure}[th!]
    \centering
    \vspace{0.5cm}
    \includegraphics[width=1\linewidth]{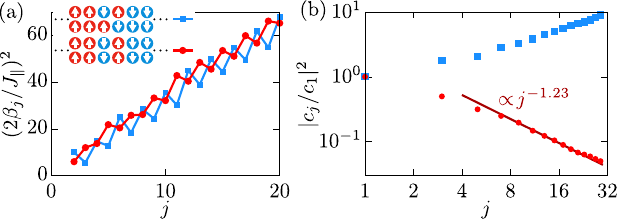}
    \vspace{-0.75cm}
    \caption{\textbf{Other initial states.} (a) Square of Lanczos coefficients for indicated initial states and $J_\perp = J_\parallel$. Note the reversal of the oscillations between the two cases. (b) Associated coefficients $|c_j|^2$ in the zero-energy eigenstate $\ket{\psi_0} = \sum_{j} c_j \ket{K_j}$.}
    \label{fig:Fig6}
    \vspace{-0.25cm}
\end{figure}

\paragraph{Appendix C: Lanczos coefficients for other initial states.--} In Fig.~\ref{fig:Fig6}, we show the Lanczos and wave function coefficients for two other indicated initial states. For the initial rung-\emph{ferromagnetic} state, the oscillations in $\beta_j$ is retained [red data in Fig.~\ref{fig:Fig6}(a)], and the wave function is localized towards the initial state [red data in Fig.~\ref{fig:Fig6}(b)]. For the state with an initial \emph{anti}ferromagnetic rung, the oscillations in $\beta_j$ reverses [blue data in Fig.~\ref{fig:Fig6}(a)], leading to a delocalized state [blue data in Fig.~\ref{fig:Fig6}(b)]. 

\end{document}


\title{Supplemental Material \\ Magnetic domains stabilized by symmetry-protected zero modes}
\maketitle
\beginsupplement
\tableofcontents

\section{Symmetry protected zero modes}
In this section, we derive a lower bound on the $E = 0$ degeneracy based on the sublattice symmetry $\hat{\mathcal{S}} = \prod_{\bi\in \mathcal{E}} (2\hat{S}^z_\bi)$, where $\mathcal{E}$ is the even sites of the lattice. \\

Note that $\hat{S}^2 = \mathbb{I}$. Hence, the Hilbert space splits into two sectors with dimensionality $N_\pm$, which are eigenvectors to $\hat{\mathcal{S}}$ with eigenvalue $\pm 1$. Based on this, and the fact that $\{\Ham, \hat{S}\} = 0$, Ref. \cite{Nicolau2026} shows that the kernel of $\Ham$ is lower bounded by the difference of the $\pm 1$ subspaces,
\begin{equation}
d(E = 0) = {\rm dim} [{\rm ker}(\Ham)] \geq |N_+(\hat{\mathcal{S}}) - N_-(\hat{\mathcal{S}})| = |{\rm tr}(\hat{\mathcal{S}})|.
\label{eq.tr_S_lower_bound}
\end{equation}
This comes about, because the sublattice symmetry enforces an even spectrum with pairs $\ket{E}, \hat{\mathcal{S}}\ket{E}$ at energy $E > 0$ and $-E < 0$. From these, an equal number of eigenstates, $(\ket{E} \pm \hat{\mathcal{S}}\ket{E})/\sqrt{2}$, to $\hat{\mathcal{S}}$ can be formed with eigenvalue $\pm  1$. Therefore, any difference between the sizes of the $\pm 1$ subspaces must come from the $E = 0$ subspace. We, thus, derive a lower bound for $d(E = 0)$ by computing ${\rm tr}(\hat{\mathcal{S}} ) = N_+(\hat{\mathcal{S}}) - N_-(\hat{\mathcal{S}})$. To this end, note that $\hat{\mathcal{S}}$ gives the parity of the number of spin-$\downarrow$ on the even sites, $\mathcal{E} = \{\bi = (x,y)\!: x + y \; {\rm even} \}$: $\hat{\mathcal{S}} = (-1)^{\hat{N}_\downarrow(\mathcal{E})}$. As a result, $N_+$ ($N_-$) equals the number of states with even (odd) $N_\downarrow(\mathcal{E})$. For fixed magnetization with $N_\downarrow$ ($N_\uparrow$) spin-$\downarrow$ (-$\uparrow$), as well as $N(\mathcal{E})$ ($N(\mathcal{O})$) even (odd) sites, this yields
\begin{align}
N_+(\hat{\mathcal{S}}) &= \sum_{N_\downarrow(\mathcal{E}) = 0,2,\dots}^{N_\downarrow} \binom{N(\mathcal{E})}{N_\downarrow(\mathcal{E})} \binom{N(\mathcal{O})}{N_\downarrow - N_\downarrow(\mathcal{E})}, \; \; N_-(\hat{\mathcal{S}}) = \sum_{N_\downarrow(\mathcal{E}) = 1,3,\dots}^{N_\downarrow} \binom{N(\mathcal{E})}{N_\downarrow(\mathcal{E})} \binom{N(\mathcal{O})}{N_\downarrow - N_\downarrow(\mathcal{E})} \Rightarrow \nn \\
{\rm tr}(\hat{S}) &= N_+(\hat{\mathcal{S}}) - N_-(\hat{\mathcal{S}}) = \sum_{N_\downarrow(\mathcal{E}) = 0}^{N_\downarrow} (-1)^{N_\downarrow(\mathcal{E})}\binom{N(\mathcal{E})}{N_\downarrow(\mathcal{E})} \binom{N(\mathcal{O})}{N_\downarrow - N_\downarrow(\mathcal{E})}
\label{eq.counting_N_pm_1}
\end{align}
Here, the factor of $\binom{N(\mathcal{E})}{N_\downarrow(\mathcal{E})}$ counts the number of ways to put $N_\downarrow(\mathcal{E})$ spin-$\downarrow$ on the $N(\mathcal{E})$ even sites, whereas $\binom{N(\mathcal{O})}{N_\downarrow - N_\downarrow(\mathcal{E})}$ gives the number of ways to position the remaining $N_\downarrow(\mathcal{O}) = N_\downarrow - N_\downarrow(\mathcal{E})$ spin-$\downarrow$ on the $N(\mathcal{O})$ odd sites. For the $N = N_x \times N_y$ square lattice, we may then divide into two cases. If $N$ is \emph{even} then $N(\mathcal{E}) = N(\mathcal{O}) = N/2$. The sum in Eq.~\eqref{eq.counting_N_pm_1} may then be computed by using two different binomial expansions. In particular,
\begin{align}
\sum_{m = 0}^{2n} \left[\sum_{k = 0}^m (-1)^k \binom{n}{k}\binom{n}{m - k}\right] x^m = (1-x)^n (1+x)^n = (1-x^2)^n = \sum_{l = 0}^{n}(-1)^l \binom{n}{l} x^{2l}.
\label{eq.binomial_expansions_1}
\end{align}
As these must match order by order in $x$, we get that for $m = 2l + 1$, $\sum_{k = 0}^m (-1)^k \binom{n}{k}\binom{n}{m - k} = 0$, whereas for even $m = 2l$, $\sum_{k = 0}^m (-1)^k \binom{n}{k}\binom{n}{m - k} = (-1)^l \binom{n}{l}$. Applying this to Eq.~\eqref{eq.counting_N_pm_1} for $n = N/2, m = N_\downarrow$, we get (for even $N$)
\begin{align}
{\rm tr}(\hat{\mathcal{S}}) = \left\{\begin{matrix}
(-1)^{N_\downarrow/2} \binom{N/2}{N_\downarrow/2}, & N_\downarrow {\; \rm even}.  \\
0, & N_\downarrow {\; \rm odd}.  \\
\end{matrix}\right. 
\label{eq.counting_N_pm_2}
\end{align}
If $N = N_x \times N_y$ is odd, then both $N_x$ and $N_y$ are odd, and $N(\mathcal{O}) = N(\mathcal{E}) - 1$. Equivalent to Eq.~\eqref{eq.binomial_expansions_1}, we then look at two different binomial expansions of $(1-x)^n(1-x)^{n-1}$. This gives
\begin{align}
\sum_{m = 0}^{2n-1} \left[\sum_{k = 0}^m (-1)^k \binom{n}{k}\binom{n-1}{m - k}\right] x^m = (1-x)^n (1+x)^{n-1} = (1-x)(1-x^2)^{n-1} = \sum_{l = 0}^{n}(-1)^l \binom{n-1}{l} (x^{2l}-x^{2l+1}).
\label{eq.binomial_expansions_2}
\end{align}
Matching this order by order in $x$ and applying it to Eq.~\eqref{eq.counting_N_pm_1} with $n = N(\mathcal{E}) = (N + 1)/2$, $N(\mathcal{O}) = N(\mathcal{E}) - 1$ and $m = N_\downarrow$ yields
\begin{align}
{\rm tr}(\hat{\mathcal{S}}) = \left\{\begin{matrix}
(-1)^{N_\downarrow/2} \binom{(N-1)/2}{N_\downarrow/2}, & N_\downarrow {\; \rm even}.  \\
(-1)^{(N_\downarrow+1)/2} \binom{(N-1)/2}{(N_\downarrow-1)/2}, & N_\downarrow {\; \rm odd}.  \\
\end{matrix}\right. 
\label{eq.counting_N_pm_3}
\end{align}
Note that these lower bound are not tight for the nearest neighbor XX model. For $N = 8\times 2$ and $N_\uparrow = N_\downarrow = 8$, we numerically find $d(E = 0) = 294$, whereas $|{\rm tr}(\hat{\mathcal{S}})| = \binom{8}{4} = 70$ from Eq.~\eqref{eq.counting_N_pm_2}.

\subsection{A tighter bound for zero magnetization}
In this subsection, we show that by combining all the identified symmetries of the XX model, we can significantly strengthen the bound derived in Eq.~\eqref{eq.counting_N_pm_2} for zero total magnetization. The given proof was discovered in a combination of our own simplifications of the problem assisted with prompts to ChatGPT-5.2.\\

To this end, we consider the combined chiral symmetry $\hat{C} = \hat{X} \hat{\mathcal{I}} \hat{\mathcal{S}}$, with the spin flip operator $\hat{X} = \prod_{\bi} (2\hat{S}^x_\bi)$, the inversion operator $\hat{\mathcal{I}} = \prod_{\bi\in \mathcal{L}} \widehat{\rm SWAP}_{\bi,\mathcal{I}(\bi)}$, and the sublattice symmetry $\hat{\mathcal{S}} = \prod_{\bi \in \mathcal{E}} (2S_\bi^z)$. Here, $\mathcal{L}$ is the set of the left half of the lattice points, $\mathcal{E}$ is the set of the even sites as above. Also, $\widehat{\rm SWAP}_{\bi,\bj}$ swaps the spins at sites $\bi$ and $\bj$, while $\mathcal{I}(\bi) = 2\bi_0 - \bi$ gives the inverted site of $\bi$ (around the lattice center $\bi_0$). Note that $[\hat{X},\Ham] = [\hat{\mathcal{I}},\Ham] = 0$. Therefore, $\hat{C}$ anticommutes with the Hamiltonian: $\{\hat{X} \hat{\mathcal{I}} \hat{S}, \Ham\} = \hat{X} \hat{\mathcal{I}} \{\hat{S}, \Ham\} = 0$. Below, we will show that ${\rm tr}(\hat{C}) = (-1)^{N/4} 2^{N/2}$ for $S^z = 0$ and even $N_x, N_y$. The same reasoning that leads to Eq.~\eqref{eq.tr_S_lower_bound}, hereby, gives a much tighter lower bound on the number of zero modes,
\begin{equation}
d(E = 0) \geq |N_+(\hat{\mathcal{C}}) - N_-(\hat{\mathcal{C}})| = |{\rm tr}(\hat{\mathcal{C}})| = \left\{\begin{matrix}
 2^{N/2}, & S^z =0 \; \& \; N_x, N_y \; {\rm even}.  \\
 0, & {\rm otherwise}.  \\
\end{matrix}\right.
\label{eq.tr_XIC}
\end{equation}
To compute ${\rm tr}(\hat{C})$, remember that the computational basis states $\ket{\sigma} = \bigotimes_\bi \ket{\sigma_\bi}$ are eigenstates of $\hat{\mathcal{S}}$ with eigenvalue $c_\sigma = (-1)^{N_\downarrow(\sigma,\mathcal{E})}$, where $N_\downarrow(\sigma,\mathcal{E})$ is the number of spin-$\downarrow$ on the even sites $\mathcal{E}$. This was also noted after Eq.~\eqref{eq.tr_S_lower_bound}. Also, note that $\hat{X}\hat{\mathcal{I}}$ maps a computational basis state to another computational basis state as: $\hat{X}\hat{\mathcal{I}}\bigotimes_\bi \ket{\sigma_\bi} = \bigotimes_\bi \ket{-\sigma_{\mathcal{I}(\bi)}}$, where $\mathcal{I}(\bi)$ is the inversion of the point $\bi$, and where we let $\sigma = \pm 1$ denote spin-$\uparrow$ and -$\downarrow$, respectively. Together, 
\begin{equation}
{\rm tr}(\hat{\mathcal{C}}) = \sum_{\sigma} \bra{\sigma} \hat{X}\hat{\mathcal{I}}\hat{\mathcal{S}}\ket{\sigma} = \sum_\sigma c_\sigma \prod_{\bi} \braket{\sigma_\bi| -\sigma_{\mathcal{I}(\bi)}}.
\end{equation}
Consequently, there is only a contribution to this trace when $\sigma_\bi = -\sigma_{\mathcal{I}(\bi)}$ for all $\bi$. This means that the contributions all reside within the zero magnetization subspace, $S^z = 0$, as every spin-$\uparrow$ necessarily has a spin-$\downarrow$ partner. It also rules out lattices with $\mathcal{I}(\bi) = \bi$. The latter happens when both $N_x,N_y$ are odd. \\

There are then two cases left to consider: either $N_x$ \emph{or} $N_y$ is odd, or both are even. In either case, $N = N_x \times N_y$ is even and the constraint $\sigma_\bi = -\sigma_{\mathcal{I}(\bi)}$ means that we can freely specify $N/2$ spins. The other $N/2$ are then specified by the constraint. This gives exactly $2^{N/2}$ possibilities. We finally need to consider the eigenvalue $c_\sigma = (-1)^{N_\downarrow(\sigma, \mathcal{E})}$ of $\hat{\mathcal{S}}$ for each spin configuration $\sigma$. When either $N_x$ or $N_y$ is odd, inversion $\hat{\mathcal{I}}$ maps the sites on the even sublattice, $\mathcal{E}$, to the odd sublattice, $\mathcal{O}$. This means that we can \emph{freely} specify the $N/2$ spins on \emph{all even} sites $\mathcal{E}$. Unlike the case considered in Eq.~\eqref{eq.counting_N_pm_2}, the $N/2$ spins on the odd lattices sites, $\mathcal{O}$, are now uniquely specified. Hence, the number of ways to get $c_\sigma = (-1)^{N_\downarrow(\sigma, \mathcal{E})} = \pm 1$ are,
\begin{align}
N_+(\hat{\mathcal{C}}) &= \sum_{N_\downarrow(\mathcal{E}) = 0,2,\dots}^{N/2} \binom{N/2}{N_\downarrow(\mathcal{E})} = 2^{N/2 - 1}, \; N_-(\hat{\mathcal{C}}) = \sum_{N_\downarrow(\mathcal{E}) = 1,3,\dots}^{N/2} \binom{N/2}{N_\downarrow(\mathcal{E})} = 2^{N/2 - 1}.
\label{eq.counting_N_pm_zero_magnetization}
\end{align}
As a result, for $N_x$ or $N_y$ odd,
\begin{equation}
{\rm tr}(\hat{\mathcal{C}}) = \sum_{\sigma} \bra{\sigma} \hat{X}\hat{\mathcal{I}}\hat{\mathcal{S}}\ket{\sigma} = N_+(\hat{\mathcal{C}}) - N_-(\hat{\mathcal{C}}) = 0.
\end{equation}
In this case, Eq.~\eqref{eq.counting_N_pm_2} gives a better bound for the number of zero modes. \\

Finally, let us consider the case where both $N_x$ \emph{and} $N_y$ are \emph{even}. In this case, inversion $\hat{\mathcal{I}}$ maps even lattice sites onto even lattice sites. As a result, we can freely specify the $N/2$ spins on the \emph{left half} of the system. The spins on the right half is then specified by the constraint $\sigma_\bi = -\sigma_{\mathcal{I}(\bi)}$. Moreover, the constraint now means that on the even lattice sites, $\mathcal{E}$, the spins always come in $(\uparrow,\downarrow)$-pairs. Hence, there are exactly $N/4$ spin-$\downarrow$ on the even sites. This means that $c_\sigma = (-1)^{N_\downarrow(\sigma,\mathcal{E})} = (-1)^{N/4}$ for all the $2^{N/2}$ states fulfilling the constraint $\sigma_\bi = -\sigma_{\mathcal{I}(\bi)}$. As a result, when both $N_x,N_y$ are even,
\begin{equation}
{\rm tr}(\hat{\mathcal{C}}) = \sum_{\sigma} \bra{\sigma} \hat{X}\hat{\mathcal{I}}\hat{\mathcal{S}}\ket{\sigma} = (-1)^{N/4} 2^{N/2}.
\end{equation}
Together, these considerations give Eq.~\eqref{eq.tr_XIC}.

\section{Exact linear scaling of Lanczos coefficients for $J_\perp = 0$.}
In this Section, we derive the linear relationship of the Lanczos coefficients: $(2\beta_{j}/J_\parallel)^2 = N_y (j - 1)$ for $J_\perp = 0$, starting from the initial state $\ket{K_1} = \ket{\Uparrow \cdots \Uparrow\Downarrow\cdots\Downarrow} = \bigotimes_{x = 0}^{N_x/2 - 1} \ket{\Uparrow}_x \bigotimes_{x = N_x/2}^{N_x - 1} \ket{\Downarrow}_x$. Here, $\ket{\Uparrow}_x$ ($\ket{\Downarrow}$ has spin-$\uparrow$ (spin-$\downarrow$) on the entire rung $x$. The given proof was discovered in a combination of our own simplifications of the problem assisted with prompts to ChatGPT-5.2. \\

For the proof, it is convenient to let $\Ham' = 2\Ham/J_\parallel$. Also, we  write the Lanczos algorithm in terms of \emph{non-normalized} wave functions. To this end, we let $\ket{\psi_0} = \ket{K_1} = \ket{\uparrow\cdots\uparrow\downarrow\cdots\downarrow}$, $\ket{\psi_1} = \Ham' \ket{\psi_{1}}$, and $\ket{\psi_{j+1}} = \Ham' \ket{\psi_j} - (\beta_j')^2\ket{\psi_{j-1}}$. Since $\alpha_j = 0$ for all $j$ [see End Matter], this gives the correct Lanczos algorithm for our current setup, with $(\beta_j')^2 = (2\beta_j / J_\parallel)^2 = \braket{\psi_{j-1}|\psi_{j-1}}/\braket{\psi_{j-2}|\psi_{j-2}}$. Hence, the problem reduces to calculating \emph{the norms of} $\ket{\psi_j}$.\\

\noindent Focus first on a single chain, $N_y = 1$. The proof then proceed in the following way.\\

\emph{1.--} At step $j$ of the Lanczos algorithm, we only need to consider a neighborhood of size $2j$ around the center of the chain. This is because the right-most spin-$\uparrow$ at most propagates $j$ steps to the right, while the left-most spin-$\downarrow$ at most propagates $j$ steps to the left. These two configurations $\ket{\uparrow\cdots\uparrow  \downarrow\cdots\downarrow\uparrow}, \ket{\downarrow\uparrow\cdots\uparrow  \downarrow\cdots\downarrow}$ of the middle $2j$ spins constitute the outer two extremes.\\

\emph{2.--} Within this region, we need to consider all possible configurations after making $j$ nearest neighbor swaps between spin-$\uparrow$ and spin-$\downarrow$, in which the spin-$\uparrow$ \emph{moves to the right} and the spin-$\downarrow$ \emph{moves to the left}. While it is clear that we need to swap the $\uparrow$- and $\downarrow$- spins, it is not immediately clear why the $\uparrow$-spins \emph{must} move to the right in each step. This follows, however, from the subtraction of $(\beta_j')^2\ket{\psi_{j-1}}$, which exactly cancels ``back propagating'' $\uparrow$-spins. \\

\emph{4.--} As a result, we must find all the \emph{integer partitions} ${\rm Par}(j)$ of $j$, i.e. the unique ways $j$ numbers can add to $j$. For example for $4$, these are $(0,0,0,4),(0,0,1,3), (0,0,2,2), (0,1,1,2), (1,1,1,1)$. This gives the possible spin configurations at step $j$ in the Lanczos algorithm. Indeed, $(0,\dots, 0, j)$ corresponds to a single spin-$\uparrow$ moving $j$ steps to the right: $\ket{\uparrow\cdots\uparrow  \downarrow\cdots\downarrow\uparrow}$. Whereas, $(1,\dots,1)$ corresponds to $j$ spin-$\uparrow$ moving 1 step to the right: $\ket{\downarrow\uparrow\cdots\uparrow  \downarrow\cdots\downarrow}$. \\

\emph{5.--} The amplitude for each spin configuration is, hereby, defined by the number of ways, $f^\lambda$, \emph{each partition occurs}. This is set by a standard Young tableau $\lambda$ defined by the partition, see Chapter 7 in Ref.~\cite{Stanley1999}. Hence, we may write
\begin{equation}
\ket{\psi_j} = \sum_{\lambda\in {\rm Par}(j)} f^\lambda \ket{\psi_j^\lambda},
\label{eq.psi_j_singlechain}
\end{equation}
in which $\ket{\psi_j^\lambda}$ is the state corresponding to the Young tableau $\lambda$. \\

\emph{6.--} It now follows from a standard identity for Young tableaux that (see Eq. (7.43) in Ref.~\cite{Stanley1999})
\begin{equation}
\braket{\psi_j|\psi_j} = \sum_{\lambda\in {\rm Par}(j)} (f^\lambda)^2 = j!.
\end{equation}

\noindent Let us finally go to the case of $N_y$ \emph{uncoupled} chains.\\

\emph{7.--} After $j$ steps, $j_y$ steps have been applied in leg $y$. The only restriction is that $\sum_{y=0}^{N_y - 1} j_y = j$. There are a total of $j!/\prod_{y = 0}^{N_y - 1} j_y!$ ways to reach a particular configuration $(j_0, j_1, \dots, j_{N_y - 1})$. The total \emph{non-normalized} state in the $j$'th step of the Lanczos algorithm for $N_y$ legs of the ladder is, thus,
\begin{equation}
\ket{\psi_j} = \sum_{\substack{j_0, \dots, j_{N_y - 1}: \\\sum_y j_y = j}} \frac{j!}{\prod_{y = 0}^{N_y - 1} j_y!} \bigotimes_{y = 0}^{N_y - 1} \ket{\psi_{j_y}(y)},
\end{equation}
where $\ket{\psi_{j_y}(y)}$ is the state in the $j_y$'th Lanczos step for the single chain $y$, found in Eq.~\eqref{eq.psi_j_singlechain}. 

\emph{8.--} It follows that the norm is
\begin{equation}
\braket{\psi_j|\psi_j} = \sum_{\substack{j_0, \dots, j_{N_y - 1}: \\\sum_y j_y = j}} \frac{\left(j!\right)^2}{\left(\prod_{y = 0}^{N_y - 1} j_y!\right)^2} \prod_{y = 0}^{N_y - 1} \braket{\psi_{j_y}(y)|\psi_{j_y}(y)} = \sum_{\substack{j_0, \dots, j_{N_y - 1}: \\\sum_y j_y = j}} \frac{\left(j!\right)^2}{\prod_{y = 0}^{N_y - 1} j_y!},
\end{equation}
using that $\braket{\psi_{j_y}(y)|\psi_{j_y}(y)} = j_y!$ from above. One may evaluate this sum by utilizing a generating function. Indeed, 
\begin{align}
&\sum_{j = 0}^\infty \frac{(N_y)^j}{j!} x^j = e^{N_y x} = \left[\sum_{j = 0}^\infty \frac{1}{j!} x^j\right]^{N_y} = \sum_{j  = 0}^\infty \Bigg[\sum_{\substack{j_0,\dots,j_{N_y-1}:\\\sum_{y=0}^{N_y-1} j_y = j}}  \frac{1}{\prod_{y = 0}^{N_y - 1} j_y!}\Bigg] x^j \Rightarrow \sum_{\substack{j_0,\dots,j_{N_y-1}:\\\sum_{y=0}^{N_y-1} j_y = j}}  \frac{1}{\prod_{y = 0}^{N_y - 1} j_y!} = \frac{(N_y)^j}{j!}.
\end{align}
The implication follows from matching the two functions order by order in $x$. It follows that: $\braket{\psi_j|\psi_j} = (N_y)^j \times j!$. \\

\emph{9.--} Finally, then, the Lanczos coefficients for $N_y$ \emph{uncoupled} legs are
\begin{align}
(\beta_j')^2 = \frac{\braket{\psi_{j-1}|\psi_{j-1}}}{\braket{\psi_{j-2}|\psi_{j-2}}} = \frac{(N_y)^{j-1} (j-1)!}{(N_y)^{j-2} (j-2)!} = N_y (j - 1).
\end{align}

This completes the proof. We have numerically checked this result for $N_y = 1,2,3$.

\section{Finite size corrections to Lanczos coefficients}
\begin{figure}[ht!]
    \centering
    \includegraphics[width=1.0\linewidth]{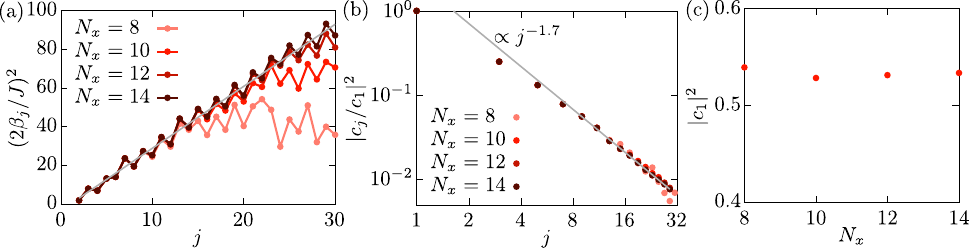}
    \vspace{-0.5cm}
    \caption{\textbf{Finite size corrections to Lanczos.} (a) Lanczos coefficients $\beta_j$ for indicated system sizes starting from the domain wall state $\ket{K_1} = \ket{\Uparrow \cdots \Uparrow \Downarrow \cdots \Downarrow}$. (b) Associated coefficients of the zero mode $\ket{\psi_0} = \sum_{{\rm odd}\; j} c_j \ket{K_j}$ showing minimal dependency on system size. The dashed gray line shows a power law fit $\sim j^{-1.7}$. (c) Associated estimate of $|c_1|^2 = |c_P|^2$ as a function of $N_x$ from Eq.~\eqref{eq.estimate_c1} using $j_0 = 11$. We use $J_\perp = J_\parallel$.}
    \label{fig:Fig8}
    \vspace{-0.25cm}
\end{figure}

In this Section, we show that the finite size corrections to the implemented Lanczos algorithm are small. To this end, we repeat the analysis in Fig. 3 of the main text starting from the domain wall state $\ket{K_1} = \ket{\Uparrow \cdots \Uparrow \Downarrow \cdots \Downarrow}$ for four lengths of the ladder: $N_x = 8, 10, 12, 14$. The results are shown in Fig.~\ref{fig:Fig8}. 

We use the results in Fig.~\ref{fig:Fig8}(b) to estimate the overlap of the initial state $\ket{K_1}$ with the zero mode $\ket{\psi_0} = \sum_j c_j \ket{K_j}$, i.e. $c_1 = c_P$. To do so, we split the sum of the norm square of the coefficients, $\sum_{{\rm odd}\; j} |c_j/c_1|^2$, into two parts. The first part is for low $j \leq j_0$, where $|c_j/c_1|^2$ does not accurately follow a power-law, whereas for the second part, we simply approximate the coefficients with the power-law fit $|c_j/c_1|^2 \simeq A j^{-\gamma}$ for $j > j_0$. In turn, $\sum_{{\rm odd}\; j} |c_j/c_1|^2 \simeq F(A,j_0,\gamma) = \sum_{{\rm odd}\; j \leq j_0} |c_j/c_1|^2 + A \sum_{{\rm odd}\; j > j_0} j^{-\gamma}$. Evaluating the power-law sum in Wolfram, we find (for $\gamma > 1$)
\begin{align}
|c_1|^{-2} &\simeq  F(A, j_0, \gamma) = \sum_{{\rm odd}\; j \leq j_0} \left|\frac{c_j}{c_1}\right|^2 + A \times \frac{(-1)^{j_0}[\zeta(\gamma,(j_0 + 1)/2) - \zeta(\gamma,j_0/2)] + 2^\gamma \zeta(\gamma,j_0)}{2^{\gamma+1}},
\label{eq.estimate_c1}
\end{align}
where $\zeta(s,a) = \sum_{n = 0}^\infty (n + a)^s$ is the Hurwitz zeta function. In Fig.~\ref{fig:Fig8}(c), we use this to estimate $|c_1|^2$ for $J_\perp = J_\parallel$ using $j_0 = 11$. This yields $|c_1|^2 \simeq 0.534$. This value is also used in Fig. 1(e) of the main text. 

\section{Exact quantum many-body scars}
In this Section, we identify two sets of exact quantum many-body scars at $E = 0$ with perfect rung-\emph{ferromagnetic} and rung-\emph{antiferromagnetic} correlations, respectively, for the $N_x \times 2$ spin ladder. These are reminiscent of recent scar states found in the spin-1 XY chain \cite{Mohapatra2026}, and are an example of Fock space cages \cite{Jonay2025}. While a single one of these states \emph{does have} an overlap with the rung-ferromagnetic states considered in the main text, this overlap decreases exponentially as $\binom{N_x}{N_x/2}^{-1}$, and is thus completely irrelevant as the system size grows for the dynamics we observe. They are, however, interesting in their own right, as they exemplify how states at the same energy have \emph{opposing} local correlations in stark constrast to the assumptions of ETH \cite{Deutsch1991}.

\subsection{Rung-ferromagnetic states}
We find that we can form states 
\begin{equation} \label{eq.RF_states}
\ket{{\rm RF}(n)} = \frac{1}{\sqrt{\mathcal{N}_n}} \frac{1}{n!}\left[\sum_{x=0}^{N_x - 1} (-1)^x \hat{S}^+_{(x,0)}\hat{S}^+_{(x,1)}\right]^n \ket{\Downarrow},
\end{equation}
where $\ket{\Downarrow} = \big |\substack{\downarrow\cdots\downarrow \\ \downarrow\cdots\downarrow}\big\rangle$. For $n = 0, \dots N_x$ these state fall into the $N_x + 1$ even magnetization sectors $S^z_n = -N_x + 2n$. They, furthermore, feature \emph{perfect} ferromagnetic correlations along all rungs of the ladder. This perfect ferromagnetic alignment on the rungs inactivates the rung XX coupling $\propto J_\perp$. Moreover, the alternating phase of $(-1)^x$ depending on the position of each rung spin-$\uparrow$ pair ensures perfect \emph{destructive interference} from terms $\propto J_\parallel$. Hence, these states sit perfectly at $E = 0$ for the pure XX model: $\Ham\ket{{\rm RF}(n)} = 0$. Moreover, the presence of ZZ rung couplings leads to an equal energy shift of all these $N_x + 1$ states to $E = N_x \Delta_\perp J_\perp / 4$ due to the perfect ferromagnetic correlations on each rung. When the ZZ coupling along the ladder becomes nonzero, $\Delta_\parallel\neq 0$, these states are no longer eigenstates of the Hamiltonian. Moreover, these rung-ferromagnetic states also remain eigenstates with zero energy in the presence of next-nearest neighbor XX couplings $\Ham_2$, as considered in Eq.~(8) of the main text.

Finally, by performing a Schmidt decomposition of $\ket{{\rm RF}(n)}$ along a cut at the centre of the ladder, we find that these states indeed show subvolume law entanglement entropy, scaling as $\sim \ln N_x$. This confirms that these states are quantum many-body scars. We proceed by showing this in detail. To this end, note that we can perform Schmidt decomposition of $\ket{{\rm RF}(n)}$ in the following manner,
\begin{align}
&\ket{{\rm RF}(n)} = \frac{1}{\sqrt{\mathcal{N}_n}}\sum_{x_1<\dots<x_n} \prod_{j=1}^n (-1)^{x_j} \hat{S}^+_{(x_j,0)}\hat{S}^+_{(x_j,1)}\ket{\Downarrow} \nn \\
&= \frac{1}{\sqrt{\mathcal{N}_n}}\sum_{l = 0}^n\left[\sum_{x_1<\dots<x_l} \prod_{j=1}^l (-1)^{x_j} \hat{S}^+_{(x_j,0)}\hat{S}^+_{(x_j,1)}\right] \times \left[\sum_{x_{l+1}<\dots<x_n} \prod_{j=l+1}^n (-1)^{x_j} \hat{S}^+_{(x_j,0)}\hat{S}^+_{(x_j,1)}\right]\ket{\Downarrow} \nn \\
&= \frac{1}{\sqrt{\mathcal{N}_n}}\sum_{l = 0}^n \sqrt{\tilde{\mathcal{N}}_{l}\tilde{\mathcal{N}}_{n-l}}\ket{u_l}_A\otimes\ket{v_l}_{\bar{A}},
\end{align}
where $l$ denotes how many spin-$\uparrow$ rungs are placed on the left half of the system and with $\tilde{\mathcal{N}}_{k} = \binom{N_x/2}{k}$. The state is, hereby, Schmidt-decomposed into the two halves of the system using the states 
\begin{align}
\!\!\!\! \ket{u_l}_A = \frac{1}{\sqrt{\tilde{\mathcal{N}}_{l}}}\left[\sum_{x_1<\dots<x_l} \prod_{j=1}^l (-1)^{x_j} \hat{S}^+_{(x_j,0)}\hat{S}^+_{(x_j,1)}\right]\ket{\Downarrow}, \ket{v_l}_{\bar{A}} = \frac{1}{\sqrt{\tilde{\mathcal{N}}_{n-l}}}\left[\sum_{x_{l+1}<\dots<x_n} \prod_{j=l+1}^l (-1)^{x_j} \hat{S}^+_{(x_j,0)}\hat{S}^+_{(x_j,1)}\right]\ket{\Downarrow}.
\end{align}
Writing $\lambda_l = \tilde{\mathcal{N}}_{l}\tilde{\mathcal{N}}_{n-l}/\mathcal{N}_n$, then gives us the von Neumann entanglement entropy for this bipartition $S_{\rm vN}(n) = -\sum_{l = 0}^n \lambda_l \ln{\lambda_l}$. Focusing on $n = N_x / 2$ and using how the binomial distribution can be approximated by the normal distribution, we get the asymptotic behavior (accurate to two decimal places for $N_x \geq 6$)
\begin{align}
S_{\rm vN}\left(\frac{N_x}{2}\right) \to \frac{1}{2}\ln{\pi N_x} - \frac{3}{2}\ln{2} + \frac{1}{2}\frac{N_x}{N_x - 1}. 
\end{align}
Hence, the entanglement entropy grows \emph{logarithmically} in the system size for these states. \\

We note, however, that only one of these rung-ferromagnetic states has an overlap with the domain wall states considered in the main text. This is the one in the $S^z = 0$ magnetization sector: $\ket{{\rm RF}(n = N_x/2)}$. Furthermore, this state is in an equal superposition of putting spin-$\uparrow$ on $N_x/2$ of the $N_x$ rungs. Therefore, its overlap is $\mathcal{N}_{N_x/2}^{-1} = \binom{N_x}{N_x/2}^{-1}$ with the domain wall product states considered in the main text. As a result, their presence does not explain the long-time stability of the domain walls. However, they do exemplify the phenomenology we found to hold more generally: in the XXZ model with arbitrary $\Delta_\perp \neq 0$, there exist quantum many-body scars which disappear as soon as $\Delta_\parallel \neq 0$. Moreover, in connection with the symmetry-protected zero modes, we stress that

\begin{enumerate}
    \item When the chiral symmetry is broken, the exponentially large degeneracy of $2^{N/2}$ in the \underline{zero magnetization} sector, $S^z = 0$, is lifted. In particular, the zero modes in the presence of the chiral symmetry are no longer degenerate \emph{anywhere} in the spectrum. 
    \item There is a \emph{special sub-sector} of $N_x + 1$ quantum many-body scars identified in the Supplemental Material. These states do not rely on the chiral symmetry. In particular, as correctly pointed out by the referee, they are still there in the presence of ZZ rung couplings, where they are all shifted from zero energy $E = 0$ to $E = N_x \Delta_\perp J_\perp / 4$, in the presence of a rung ZZ coupling, $\Delta_\perp J_\perp$. Note that these scar states are the only states that undergoes this particular shift in energy.
    \item Importantly, these $N_x + 1$ states belong to \emph{distinct} magnetization sectors. Hence, the constant shift $E = N_x \Delta_\perp J_\perp / 4$ of these states \emph{does not} lead to any degeneracies at a \emph{fixed total magnetization}, $S^z$. It is only when one looks across all magnetization sectors, that this leads to an $N_x + 1$ fold degeneracy.
\end{enumerate}

We also note that these scars are similar to but not entirely equivalent with a \emph{spectral generating algebra} (SGA) \cite{Chandran2023} or dynamical symmetry \cite{Tindall2020}, for which one can find an dynamical symmetry operator $\hat{Q}$ such that $[\Ham,\hat{Q}] = \omega \hat{Q}$. In such cases, for a state $\ket{\psi_0}$ with energy $E_0$ and $\hat{Q}\ket{\psi_0} \neq 0$, it follows that $\Ham\hat{Q}^n\ket{\psi_0} = (E_0 + n\omega)\ket{\psi_0}$ [since $[\Ham, \hat{Q}^n] = n\omega \hat{Q}^n$. Based on Eq.~\eqref{eq.RF_states} it is therefore tempting to set $\hat{Q} = \sum_{x=0}^{N_x - 1} (-1)^x \hat{S}^+_{(x,0)}\hat{S}^+_{(x,1)}$. However, this operator is \emph{not} a dynamical symmetry of the XX Hamiltonian. Instead, we can form an entire series of operators, $\hat{Q}_n$ such that $[\Ham,\hat{Q}_n] = 0$. Here, 
\begin{equation} \label{eq.Qn_s}
\hat{Q}_n = \left[\sum_{x=0}^{N_x - 1} (-1)^x \hat{S}^+_{(x,0)}\hat{S}^+_{(x,1)}\right]^n \hat{P}_{\Downarrow},
\end{equation}
with the projection onto the all-down state $\hat{P}_{\Downarrow} = \ket{\Downarrow}\bra{\Downarrow}$. As a result of the all-$\downarrow$ projection, each symmetry $\hat{Q}_n$ leads to a single scar state, $\ket{\rm RF(n)}$.

\subsection{Rung-antiferromagnetic states}
It is also worth pointing out, that on top of these ferromagnetic states, we also find scars that feature perfect rung-\emph{antiferromagnetic} states,
\begin{align} \label{eq.rung_antiferromagetic_states}
\ket{{\rm RA_e}} = \frac{1}{\sqrt{\mathcal{N}}} \sum_{{\rm even\;} n} &  \frac{1}{n!} \left[\sum_{x=0}^{N_x - 1} (-1)^x \hat{S}^+_{(x,0)}\hat{S}^-_{(x,1)}\right]^{ n}  \Big|\substack{\uparrow\dots\uparrow \\ \downarrow\dots\downarrow}\Big\rangle,     \nn \\
\ket{{\rm RA_o}} = \frac{1}{\sqrt{\mathcal{N}}}  \sum_{{\rm odd\;} n} &  \frac{1}{n!} \left[\sum_{x=0}^{N_x - 1} (-1)^x \hat{S}^+_{(x,0)}\hat{S}^-_{(x,1)}\right]^{ n} \Big|\substack{\uparrow\dots\uparrow \\ \downarrow\dots\downarrow}\Big\rangle,    
\end{align}
with exactly opposite energy $E = -N_x \Delta_\perp J_\perp / 4$ to the states $\ket{{\rm RF}(n)}$ in Eq.~\eqref{eq.RF_states} for $\Delta_\parallel = 0$. This construction works \emph{only} in the $S^z = 0$ magnetization sector and only gives rise to these two states. Here, $\mathcal{N} = 2^{N_x - 1}$. As a result, when $\Delta_\perp = 0$, there are states at the same energy, $E = 0$, with maximally different local correlations, thereby strongly deviating from the assumptions of ETH.

\section{Towards two dimensions} 
\begin{figure}[ht!]
    \centering
    \includegraphics[width=0.6\linewidth]{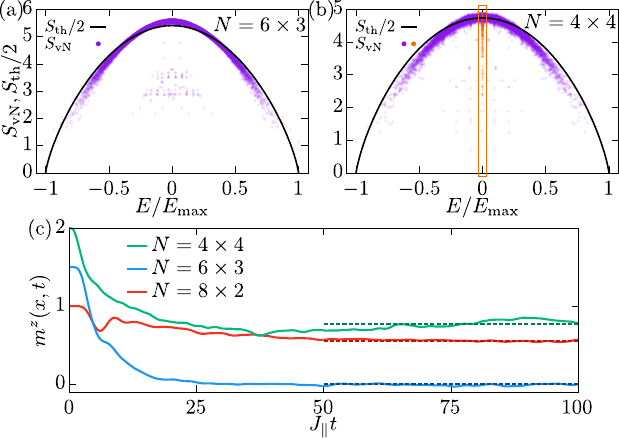}
    \vspace{-0.5cm}
    \caption{\textbf{Towards two dimensions.} (a), (b) $S_{\rm vN}(E)$ vs energy for indicated system sizes in the $S^z = 0$ magnetization sector for the pure XX model ($\Delta_\perp = \Delta_\parallel = 0$). This is compared to half the thermal entropy $S_{\rm th}(\beta = E^{-1})/2$. There is again a strong even/odd effect: for $N_y = 3$ legs (a), there is no zero mode degeneracy, whereas a large degeneracy reappears for $N_y = 4$ in (b) [orange box]. (c) The resulting magnetization dynamics starting from a product state with all spin-$\uparrow$ on the left in turn features an even/odd effect: for an even number of legs, $N_y = 2,4$, the magnetization reaches a nonzero steady state value (dashed lines). Parameters: for $N_y = 2,3$, we use $J_\perp = J_\parallel$. For $N_y = 4$, we use $J_\perp = 2J_\parallel$.}
    \label{fig:Fig7}
    \vspace{-0.25cm}
\end{figure}

In this Section, we provide numerical evidence that the non-ergodic behavior is not specific to the $N = N_x \times 2$ spin ladder. Figure \ref{fig:Fig7}, hereby, considers $N = 6\times 3, 4\times 4$ geometries of the lattice. The entanglement entropies computed in Figs.~\ref{fig:Fig7}(a) and \ref{fig:Fig7}(b) show that scar states remain in these cases, again with a striking even/odd effect. Indeed, the $E = 0$ degeneracy vanishes completely for $N_y = 3$ and $N_\uparrow = N/2$, whereas it jumps back up to $d = 306$ states for the $4\times 4$ square. This also manifests in the magnetization dynamics, in which an initial domain wall state with the $N/2$ spin-$\uparrow$ on the left retains a stable magnetization profile only for the even cases $N_y = 2,4$, whereas for $N_y = 3$ the system thermalizes. This demonstrates that our findings are not special cases limited to the spin ladder. We note, however, that to retain a magnetization at the same level as for the spin ladder for $N_y = 4$, we increase the rung coupling to $J_\perp = 2J_\parallel$. This suggests that the magnetic domains will slowly fade as the \emph{two-dimensional} thermodynamic limit is approached: $N_x,N_y\to \infty$.

\section{Matrix elements at nonzero energies}
In this section, we give further details on why the second term in Eq. (6) of the main text, 
\begin{align}
\delta \bar{S}^z_\bi = \sum_{\substack{n,m: \\ E_n = E_m \neq 0} } c_n^* c_m \bra{n} \hat{S}^z_\bi \ket{m},
\label{eq.temporal_average_fluctuations_term}
\end{align}
contributes extremely little to $\bar{S}^z_\bi$. To this end, note that the spin flip operator $\hat{X} = \prod_{\bf i} (2\hat{S}^x_{\bf i})$ commutes with the Hamiltonian. Therefore, we can form simultaneous eigenstates $\ket{n}$ to both $\hat{X}$ (with eigenvalue $x_n$) and $\hat{H}$ (with energy eigenvalue $E_n$). For any two such eigenstates $\ket{n},\ket{m}$, it follows that
\begin{align}
x_n \bra{n} \hat{S}^z_\bi \ket{m} = \bra{n} \hat{X} \hat{S}^z_\bi \ket{m} = - \bra{n}  \hat{S}^z_\bi \hat{X}\ket{m} = -x_m  \bra{n} \hat{S}^z_\bi \ket{m}. 
\end{align}
Here, we use that $\hat{X}$ anticommutes with $\hat{S}^z_\bi$. It follows that either $x_n = -x_m$ \emph{or} $\bra{n} \hat{S}^z_\bi \ket{m} = 0$. This shows that in this basis all expectation values vanish: $\bra{n} \hat{S}^z_\bi \ket{n} = 0$. So the only possible contributions from nonzero energy states are from \emph{degenerate} subspaces, $E_n = E_m$ with $n \neq m$ and with the additional requirement that $x_n = -x_m$. Furthermore, to contribute to Eq.~\eqref{eq.temporal_average_fluctuations_term}, the overlaps $c_n = \braket{n|\Psi(0)},c_m = \braket{m|\Psi(0)}$ must also be non-vanishing. This combination of nonzero matrix elements as well as nonzero $c_n, c_m$ turns out to be exceedingly rare. In particular, for $N_x = 8$ and $J_\perp = J_\parallel$, having a total of $\binom{16}{8} = 12\, 870$ states at zero total magnetization $S^z = 0$, there are only \emph{two} nonzero contributions to Eq.~\eqref{eq.temporal_average_fluctuations_term}. These come from $E_n = E_m \simeq \pm 1.174 J_\parallel$, and only amounts to $c_n^* c_m \bra{n} \hat{S}^z_\bi \ket{m} \sim 10^{-6}$. 

\section{Free fermions}
In this Section, we show that for free fermions with nearest neighbor hoppings $J_\parallel/2, J_\perp/2$, the rung-ferromagnetic initial states behave exactly like a collection of $N_y$ uncoupled one-dimensional chains. Hence, for free fermions there is no dependency on $J_\perp$ for these initial states in extreme contrast to the very strong dependency for the XX model, i.e. for hard-core bosons. \\

We start from the nearest neighbor tight binding model of free fermions,
\begin{align}
\Ham &= \sum_{\substack{\lambda = \parallel,\perp \\ \braket{\bi,\bj}_\lambda}} \frac{J_\lambda}{2} \left[\hat{f}^\dagger_\bi\hat{f}_\bj + \hat{f}^\dagger_\bi\hat{f}_\bj\right],
\label{eq.H_fermions}
\end{align}
Next, we diagonalize Eq.~\eqref{eq.H_fermions} by decomposing the real space fermionic operators in terms of crystal momenta modes (with hard-wall boundary conditions as considered in the main text for the spin model)
\begin{align}
\hat{f}^\dagger_\bi = \frac{2}{\sqrt{(N_x + 1)(N_y + 1)}} \sum_{\bk} \sin(k_x x) \sin(k_y y) \hat{f}^\dagger_\bk,
\label{eq.f_k}
\end{align}
with $\bk = (k_x, k_y)$, fulfilling the quantization condition $k_j = n_j\pi / (N_j + 1)$ for $n_j = 1, \dots, N_j$ and $j = x,y$. Inserting this into Eq.~\eqref{eq.H_fermions} yields
\begin{align}
\Ham &= \sum_{\bk} \varepsilon_\bk \hat{f}^\dagger_\bk \hat{f}_\bk,
\label{eq.H0_fermions_momentum_space}
\end{align}
with $\varepsilon_\bk = \varepsilon_\parallel(k_x) + \varepsilon_\perp(k_y) = J_\parallel \cos(k_x) + J_\perp \cos(k_y)$. Suppose now that we start from a product state, $\ket{\Psi(0)}$ with $N_f = n N_y$ fermions, for some integer $n = 1, \dots N_x$. Suppose further, that the fermions fill up exactly $n$ columns of the rectangular lattice, call this set of columns $C_x$. In this manner, there are $n$ fully occupied columns and $N_x - n$ empty columns initially. We may write
\begin{equation}
\ket{\Psi(0)} = \prod_{x\in C_x} \prod_{y = 0}^{N_y - 1} \hat{f}^\dagger_{(x,y)} \ket{0}.
\label{eq.column_filled_fermion_state}
\end{equation}
Importantly, because each column is fully occupied in real space, the fermionic nature of the particles entail that the state is also fully occupied along $k_y$. I.e., we may use that $\prod_{y = 0}^{N_y-1} \hat{f}^\dagger_{(x,y)} \ket{0} = \prod_{k_y} \hat{f}^\dagger_{(x,k_y)}\ket{0}$, with the partially Fourier transformed modes $\hat{f}^\dagger_{(x,k_y)} = \sqrt{2/(N_y+1)} \sum_{y} \sin(k_y y) \hat{f}^\dagger_{(x,y)}$. Importantly, these modes diagonalize the free-fermion Hamiltonian along the $y$-direction, 
\begin{align}
\Ham =  \sum_{(x, k_y)} \Big[ \frac{J_\parallel}{2} \left(\hat{f}^\dagger_{(x + 1, k_y)} \hat{f}_{(x, k_y)} + {\rm H.c.}\right)  +  \varepsilon_\perp(k_y) \hat{f}^\dagger_{(x, k_y)} \hat{f}_{(x, k_y)}\Big].
\label{eq.H0_fermions_partially_transformed}
\end{align}
In particular, the operators
\begin{align}
\hat{N}_{k_y} = \sum_{x} \hat{f}^\dagger_{(x,k_y)}\hat{f}_{(x,k_y)}
\label{eq.symmetries}
\end{align}
are conserved quantities, $[\hat{N}_{k_y}, \Ham] = 0$, such that the Hamiltonian can be written as two commuting terms $\Ham = \Ham_{\parallel} + \Ham_{\perp}$ with $\Ham_{\perp} = \sum_{k_y} \varepsilon_\perp(k_y) \hat{N}_{k_y}$. Finally, $\ket{\Psi(0)}$ is an eigenstate of $\Ham_{\perp}$ with vanishing eigenvalue,
\begin{align}
\Ham_{\perp} \ket{\Psi(0)} = n \sum_{k_y} \varepsilon_\perp(k_y) \ket{\Psi(0)} = 0,
\end{align}
using that the number of fermions in state $k_y$ is exactly $N_{k_y} = n$ (the number of filled rungs) and that $\sum_{k_y} \varepsilon_\perp(k_y) = 0$. Since $[\Ham_{\perp},\Ham_{\parallel}] = 0$, the time-evolved state is 
\begin{align}
\ket{\Psi(t)} = e^{-i\Ham_{\parallel} t}e^{-i\Ham_{\perp} t} \ket{\Psi(0)} = e^{-i\Ham_{\parallel} t} \ket{\Psi(0)}.
\end{align}
Moreover, $\Ham_{\parallel} = \sum_{y} \Ham_{\parallel}(y)$ is a sum of $N_y$ commuting terms. Therefore, 
\begin{align}
\ket{\Psi(t)} &= \bigotimes_{y = 0}^{N_y - 1} e^{-i\Ham_{\parallel}(y) t} \ket{\Psi_y(0)},
\end{align}
using that the initial state is a product state of the $N_y$ columns of the lattice. As a result, the time-evolved state remains a product state and the only dependence of the coupling along the columns of the lattice, $J_\perp$, is in the overall phase factor in front. This means that the free fermion model initialized in one of the column-filled states of Eq.~\eqref{eq.column_filled_fermion_state} gives rise to dynamics that is completely independent of the number of rows of the lattice, i.e. the whole $N_x\times N_y$ rectangular lattice acts as a one-dimensional chain of length $N_x$.

\bibliographystyle{apsrev4-2}
\bibliography{bibliography}